\documentclass[aps,prb,twocolumn,superscriptaddress]{revtex4-2}

\usepackage[dvipdfmx]{graphicx}% Include figure files
\usepackage{dcolumn}% Align table columns on decimal point
\usepackage{bm}% bold math
\usepackage{amsmath}

\begin{document}

%\preprint{}

\title{Spin and charge dynamics of doped one-dimensional Mott insulators}


\author{Takami Tohyama}
\email[]{tohyama@rs.tus.ac.jp}
%\thanks{}
\affiliation{Department of Applied Physics, Tokyo University of Science, Katsushika, Tokyo 125-8585, Japan}

\author{Hiroki Maeda}
%\email[]{}
%\thanks{}
\affiliation{Department of Applied Physics, Tokyo University of Science, Katsushika, Tokyo 125-8585, Japan}

\author{Kenji Tsutsui}
%\email{tsutsui.kenji@qst.go.jp}
\affiliation{Synchrotron Radiation Research Center, National Institutes for Quantum Science and Technology, Hyogo 679-5148, Japan}

\author{Shigetoshi Sota}
\affiliation{Computational Materials Science Research Team, RIKEN Center for Computational Science (R-CCS), Kobe, Hyogo 650-0047, Japan}
\affiliation{Quantum Computational Science Research Team, RIKEN Center for Quantum Computing (RQC), Wako, Saitama 351-0198, Japan}

\author{Seiji Yunoki}
\affiliation{Computational Materials Science Research Team, RIKEN Center for Computational Science (R-CCS), Kobe, Hyogo 650-0047, Japan}
\affiliation{Quantum Computational Science Research Team, RIKEN Center for Quantum Computing (RQC), Wako, Saitama 351-0198, Japan}
\affiliation{Computational Quantum Matter Research Team, RIKEN Center for Emergent Matter Science (CEMS), Wako, Saitama 351-0198, Japan}
\affiliation{Computational Condensed Matter Physics Laboratory, RIKEN Cluster for Pioneering Research (CPR), Saitama 351-0198, Japan}

\date{\today}

\begin{abstract}
Recent angle-resolved photoemission spectroscopy (ARPES) for a newly discovered doped one-dimensional (1D) Mott insulator Ba$_{2-x}$Sr$_x$CuO$_{3+\delta}$ proposed the presence of attractive charge interactions between neighboring sites if a 1D Hubbard model is employed to describe the material. On the other hand, long-range repulsive charge interactions were proposed to be crucial for understanding the optical excitations of undoped 1D Mott insulators. Motivated by these contrasting proposals, we perform density-matrix renormalization group calculations of the single-particle spectral function as well as dynamical spin and charge structure factors for two kinds of extended 1D Hubbard models. The spectral function for the long-range repulsive model less agrees with the ARPES data as compared with that for the attractive one. The dynamical spin and charge structure factors also exhibit contrasting behaviors between the two models: The attractive model shows less momentum dependence of the integrated weight in the dynamical spin structure factor and more accumulation of the weight on the upper edge of particle-hole-type excitation in the dynamical charge structure factor.  These features give a good criterion of an appropriate model for doped 1D Mott insulators, which can be judged by resonant-inelastic x-ray scattering experiments.

\end{abstract}

%\keywords{}

\maketitle

\section{Introduction}
\label{Sec1}

The introduction of carriers into Mott insulators gives novel quantum phenomena due to strong electron correlation. Such systems are called doped Mott insulators. Cuprate high-temperature superconductors are typical of two-dimensional doped Mott insulators, where chemical substitution introduces carriers into the CuO$_2$ planes. In contrast, doping carriers into one-dimensional (1D) Mott insulators via chemical substitution was unsuccessful. However, recent thin film technology has successfully synthesized doped 1D Mott insulators Ba$_{2-x}$Sr$_x$CuO$_{3+\delta}$~\cite{Chen2021}, where a 1D Cu-O chain gives a conducting path for hole carriers. 

In undoped 1D Mott insulators such as SrCuO$_2$ and Sr$_2$CuO$_3$, angle-resolved photoemission spectroscopy (ARPES) has shown dispersive branches composed of spinons and holons whose lowest binding energy is located at momentum $k=\pi/2$~\cite{Kim1996,Kim1997,Fujisawa1999,Kim2006}. In these ARPES experiments, a single hole is introduced  into the lower-Hubbard band (more precisely the Zhang-Rice singlet band) by removing an electron by the photoemission process. The presence of spinon and holon branches is a consequence of the separation of spin and charge degrees of freedom, i.e., spin-charge separation, inherent in 1D correlated systems~\cite{Essler2005}.
      
The ARPES spectrum for hole-doped Ba$_{2-x}$Sr$_x$CuO$_{3+\delta}$~\cite{Chen2021} has also exhibited dispersive branches corresponding to spinons and holons for momentum $k<k_\mathrm{F}$, where $k_\mathrm{F}$ is the Fermi momentum. For $k>k_\mathrm{F}$, the branch starting from $k=3k_\mathrm{F}$, which we call the 3$k_\mathrm{F}$ branch, is expected to appear according to theoretical calculations of the electron-removal single-particle spectral function in the 1D Hubbard model~\cite{Benthien2004}. However, the 3$k_\mathrm{F}$ branch has not been clearly detected around the binding energy of 0.7~eV. In contrast, another branch corresponding to the folding of the holon branch with respect to $k=k_\mathrm{F}$, which was called the secondary holon band in Ref.~\cite{Benthien2004}, has been observed though its spectral weight is weak~\cite{Chen2021}. The spectral weight of this holon-folding branch was proposed to be enhanced if a nearest-neighbor attractive charge interaction is introduced into the 1D Hubbard model~\cite{Chen2021}. In fact, the calculated spectral functions have reproduced very well the ARPES data~\cite{Chen2021,Feiguin2024,Wang2024}. The origin of the attractive interaction has been attributed to nonlocal electron-phonon interactions~\cite{Wang2021,Tang2023}.

In an organic undoped 1D Mott insulator ET-F$_2$TCNQ, pump-probe spectroscopy has revealed the presence of  a biexciton consisting of a pair of holon-doublon bound states~\cite{Miyamoto2019}, which is located just below a Mott-gap absorption peak. Long-range repulsive C
oulomb interactions are necessary for forming the biexciton. The energy difference between the biexciton and the Mott-gap peak is related to the third-neighbor Coulomb interaction. 

Then, arising questions are (i) whether the long-range repulsive interactions explain ARPES data in Ba$_{2-x}$Sr$_x$CuO$_{3+\delta}$, although the nonlocal electron phonon interactions may weaken the repulsive interaction~\cite{Wang2021}, and (ii) what are the differences of spin and charge dynamics between the attractive and long-range repulsive 1D Hubbard models.

In this paper, to give hints for these questions, we first perform a density-matrix renormalization group (DMRG) calculation of the single-particle spectral function for extended Hubbard models. For a negative nearest-neighbor interaction, an enhancement of the holon-folding spectral weight is observed around the binding energy of 0.6~eV as was reported before~\cite{Chen2021}. For long-range repulsive interactions, the spectral weight at the same binding energy is located in a momentum region lower than that expected from the $3k_\mathrm{F}$ branch, which is consistent with the negative case, but its agreement with the ARPES data in Ba$_{2-x}$Sr$_x$CuO$_{3+\delta}$ is less satisfactory. We next examine dynamical spin and charge structure factors for the same models. We find that the integrated weight of the dynamical spin structure factor up to $\sim 0.6$~eV becomes less momentum dependent for $q>2k_\mathrm{F}$ in the case of an attractive interaction than in the case of long-range repulsive interactions. Such a different momentum dependence is a good criterion for judging whether an attractive or repulsive interaction is appropriate for doped 1D Mott insulators. In the dynamical charge structure factor, we find that the spectral weight almost concentrates on the upper edge of the particle-hole-type excitation for the attractive interaction, in contrast to the case of repulsive interactions where a low-energy $4k_\mathrm{F}$ excitation significantly contributes to the spectral weight. This is also a good criterion to judge an appropriate model for doped 1D Mott insulators. We propose that resonant-inelastic x-ray scattering (RIXS) experiments will be a useful tool to detect differences in the spin and charge dynamics that come from the difference between attractive and repulsive interactions and can determine the model appropriate for the doped 1D cuprates.

This paper is organized as follows. The 1D extended Hubbard model and dynamical DMRG method are introduced in
Sec.~\ref{Sec2}. In Sec.~\ref{Sec3}, calculated results of the single-particle spectral function as well as dynamical spin and charge structure factors are shown for the extended Hubbard model with negative and repulsive interactions as well as the pure Hubbard model. Finally, a summary is given in Sec.~\ref{Sec4}.

\section{Model and method}
\label{Sec2}
We investigate an extended one-dimensional (1D) Hubbard model,
\begin{align}
H=&-t\sum_{i,\sigma}\left(c^{\dagger}_{i,\sigma} c_{i+1,\sigma}+\text{H.c.}\right)+U\sum_{i}n_{i,\uparrow}n_{i,\downarrow}\notag\\
  &+\sum_{i}\sum_{m=1}^3 V_m n_{i}n_{i+m},
\label{H}
\end{align} 
where $c^{\dagger}_{i,\sigma}$ ($c_{i,\sigma}$) is the creation (annihilation) operator of an electron with spin $\sigma$ at site $i$, $n_i= n_{i,\uparrow}+ n_{i,\downarrow}$ with $n_{i,\sigma}=c^{\dagger}_{i,\sigma}c_{i,\sigma}$, $t$ is the hopping constant, $U$ is on-site repulsive Coulomb interaction, and $V_m$ ($m$=1, 2, and 3) is the $m$th neighbor interaction. We take $U/t=8$ proposed for Ba$_{2-x}$Sr$_x$CuO$_{3+\delta}$ with $t=0.6$~eV~\cite{Chen2021}. In the following, we set $t$ as an energy unit ($t=1$) and the reduced Planck constant $\hbar=1$. The lattice constant is also taken to be unity. The hole concentration in doped 1D Mott insulator is represented by $x$, which is defined as $x=1-N_\mathrm{e}/N$, where $N_\mathrm{e}$ is the electron number in an $N$-site lattice. 

We examine three momentum-dependent dynamical quantities: the dynamical charge structure factor $N(q,\omega)$, dynamical spin structure factor $S(q,\omega)$, and spectral function $A(k,\omega)$. Using the ground state of an $N_\mathrm{e}$-electron system, $\left|N_\mathrm{e}\right\rangle$, with energy $E_\mathrm{GS}$, the dynamical quantities are given by
\begin{flalign}
&N(q,\omega)=-\frac{1}{N\pi} \mathrm{Im} \left\langle N_\mathrm{e} \right| \tilde{N}_{-q} \frac{1}{\omega  - H + E_\mathrm{GS}+i\gamma } \tilde{N}_q \left| N_\mathrm{e} \right\rangle \label{Nqw},&\\
&S(q,\omega)=-\frac{1}{N\pi} \mathrm{Im} \left\langle N_\mathrm{e} \right| S_{-q}^z \frac{1}{\omega  - H + E_\mathrm{GS}+i\gamma } S_q^z \left| N_\mathrm{e} \right\rangle \label{Sqw},&
\end{flalign}
and
$A(k,\omega)=A_-(k,\omega)+A_+(k,\omega)$
with
\begin{align}
&A_\pm(k,\omega)=\notag\\
&-\frac{1}{N\pi} \mathrm{Im}\sum_\sigma\left\langle N_\mathrm{e} \right| a_{k,\sigma}^\dagger \frac{1}{\omega  \mp (H - E_\mathrm{GS})-\mu+i\gamma } a_{k,\sigma} \left| N_\mathrm{e} \right\rangle \label{Akw},
\end{align}
where $\tilde{N}_q=N_q-\left\langle N_\mathrm{e}\right| N_q \left| N_\mathrm{e}\right\rangle$, the operator $N_q$ ($S_q^z$) is the Fourier component of $n_i$ [$S_i^z=(n_{i,\uparrow}-n_{i,\downarrow})/2$], $\mu$ is the chemical potential, and the Lorentzian broadening factor $\gamma$ is a small positive number. $A_+$ ($A_-$) in Eq.~(\ref{Akw}) is the electron-addition (electron-removal) spectral function and $a_{k,\sigma}=c_{k,\sigma}^\dagger$ ($c_{k,\sigma}$) for $A_+$ ($A_-$), which is the Fourier component of $c_{i,\sigma}^\dagger$ ($c_{i,\sigma}$).

\begin{figure*}[t]
\includegraphics[width=18cm]{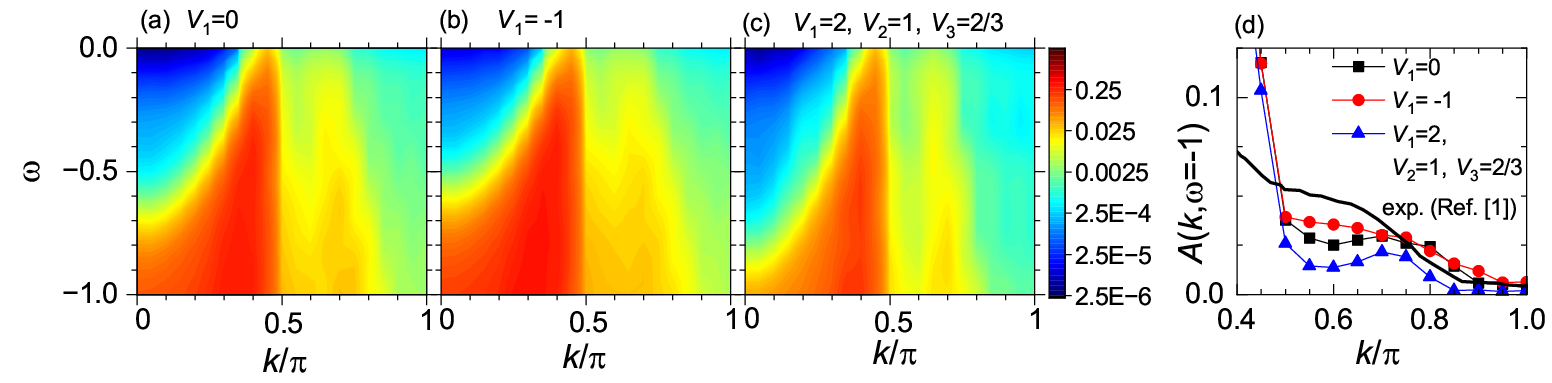}% Here is how to import EPS art
\caption{\label{fig1}Electron-removal spectral function $A_-(k,\omega)$ for an $N=40$ extended Hubbard model with $U=8$ at hole concentration $x=0.1$. (a) The Hubbard model ($V_1=0$), (b) the negative $V$ model ($V_1=-1$), and (c) the long-range $V$ model ($V_1=2$, $V_2=1$, and $V_3=2/3$). The logarithmic scale of the spectral weight is shown in the bar at the right-hand side of (c). (d) Spectral weight at $\omega=-1$ from $k=0.4\pi$ to $\pi$ for the three models. The solid curve represents the experimental ARPES weight at $x=0.09$ taken from Ref.~\cite{Chen2021}.}
\end{figure*}

We calculate Eqs.~(\ref{Nqw}) and (\ref{Sqw}) for an $N=80$ periodic lattice and Eq.~(\ref{Akw}) for an $N=40$ periodic lattice using dynamical DMRG~\cite{Jeckelmann2002}, where we use three kinds of target states: for example, for $N(q,\omega)$, (i) $\left| N_\mathrm{e}\right\rangle$, (ii) $\tilde{N}_q \left| N_\mathrm{e} \right\rangle$, and (iii) $(\omega - H + E_\mathrm{GS}+i\gamma)^{-1} \tilde{N}_q \left| N_\mathrm{e} \right\rangle$. The target state (iii) is evaluated using a kernel-polynomial expansion method~\cite{Sota2010}, where the Lorentzian broadening $\gamma$ is replaced by Gaussian broadening with a half width at half maximum of 0.04 (0.1) for $N(q,\omega)$ and $S(q,\omega)$ [$A(k,\omega)$]. In our numerical calculations, we divide the energy interval $[0,E]$ by $\lambda$-mesh points and target all of the points at once: $(E,\lambda)=(2,100)$, $(1,50)$, and $(-1.7, 34)$ for $N(q,\omega)$, $S(q,\omega)$, and $A(k,\omega)$, respectively. To perform the dynamical DMRG, we use the truncation number $m=4000$ for $N(q,\omega)$ and $S(q,\omega)$ and $m=2000$ for $A(k,\omega)$.

The chemical potential $\mu$ is defined as an energy where the density of states from the electron-removal spectral function $N_-(\omega)=\sum_kA_-(k,\omega)$ is equal to that from the electron-addition spectral function $N_+(\omega)=\sum_kA_+(k,\omega)$. 

\section{Results}
\label{Sec3}

We examine three parameter sets of the extended Hubbard model in Eq.~(\ref{H}) with $x=0.1$ and 0.2: (i) $V_1=V_2=V_3=0$ denoted as $V_1=0$ or the Hubbard model in the rest of this paper, (ii) $V_1=-1$, $V_2=V_3=0$ (proposed in Ref.~\cite{Chen2021}) denoted as $V_1=-1$ or the negative $V$ model, and (iii) $V_1=2$, $V_2=1$, and $V_3=2/3$ (proposed in Ref.~\cite{Miyamoto2019})  denoted as the long-range $V$ model. 

The choice of these parameters is based on experimental facts. For the hole concentration $x$, we refer to the fact that the holon-folding branch is realized as long as $x\lesssim 0.2$~\cite{Chen2021}. Therefore, we choose the two values, $x=0.1$ and 0.2. The Hubbard model [the parameter set (i)] is introduced as a reference mode for the other parameter sets, (ii) and (iii). For the parameter set (ii), we take $V_1=-1$ from Ref.~\cite{Chen2021} as a realistic value of the negative $V$. In the long-range $V$ model, the ratio of $V_1$, $V_2$ and $V_3$ to $t$ is taken from Ref.~\cite{Miyamoto2019}, where the biexciton appears just below the Mott-gap edge. The value of $V_1$ that is nearly twice  $t$ has been reported in ET-F$_2$TCNQ~\cite{Miyamoto2019} as well as the cuprate 1D Mott insulator Sr$_2$CuO$_3$~\cite{Sota2010,Ono2004}. We note that in 1D Mott insulators, observed optical conductivities show a peak structure at the Mott-gap edge. The peak has been assigned as a holon-doublon (H-D) excitonic peak. We emphasize that $V_1$ is more than $2t$ when a H-D bound state appears~\cite{Stephan1996}.

\subsection{Spectral function}
Figure~\ref{fig1} shows the electron-removal spectral function at $x=0.1$ on an $N=40$ Hubbard lattice for the three models. Since the Fermi momentum $k_\mathrm{F}=\pi(1-x)/2$ in the 1D Hubbard model, $k_\mathrm{F}=0.45\pi$ at $x=0.1$ and thus $3k_\mathrm{F}=2\pi-3\times k_\mathrm{F}=0.65\pi$.

In the Hubbard model ($V_1=0$) shown in Fig.~\ref{fig1}(a), a dispersion traced by the maximum intensity for each $\omega$ is the holon branch whose momentum at $\omega=-1$ is around $k=0.35\pi$,  where the momentum shift from $k_\mathrm{F}$ is $k_\mathrm{shift}=k_\mathrm{F}-0.35\pi=0.1\pi$. Note that $\omega=-1$ corresponds to $\omega\sim -0.6$~eV, which is close to the binding energy 0.7~eV examined in Ref.~\cite{Chen2021}.
%Accordingly, 
At $\omega=-1$, the $3k_\mathrm{F}$ and holon-folding branches are expected to be located at $k=3k_\mathrm{F}+k_\mathrm{shift}=0.75\pi$ and $k=k_\mathrm{F}+k_\mathrm{shift}=0.55\pi$, respectively. 
In fact, we find in Fig.~\ref{fig1}(a) a weak and broad weight centered at $k=0.75\pi$ for $\omega=-1$, which can be assigned as the $3k_\mathrm{F}$ branch, and there is a small dip around  $k=0.6\pi$. 

In the negative $V$ model shown in Fig.~\ref{fig1}(b), there is no dip around $k=0.6\pi$ at $\omega=-1$ unlike the Hubbard model but weights are broadly distributed from $k=0.5\pi$ to $k=0.7\pi$, which is an indication of the presence of the holon-folding branch. We note that our preliminary results indicate the presence of a peak structure around $k=0.6\pi$ if we include long-range attractive interactions up to the third neighbors (not shown here). In the long-range $V$ model shown in Fig.~\ref{fig1}(c), a peak appears around $k=0.7\pi$ at $\omega=-1$. 

We summarize the $k$ dependence of spectral weight above $k=0.4\pi$ for the three models in Fig.~\ref{fig1}(d), together with the ARPES data at $x=0.09$ taken from Ref.~\cite{Chen2021}. We can clearly see that a broad peak position of the Hubbard model (black squares) shifts to a lower momentum around $k=0.6\pi$ with introducing negative $V$ (red circles), resulting in no dip structure at $k=0.6\pi$, which is similar to  the experimental data (solid curve). This is consistent with previous reports~\cite{Chen2021,Feiguin2024,Wang2024}. It is interesting to notice that the long-range $V$ model (blue triangles) is just between the Hubbard model and negative $V$ model as evidenced by the presence of a peak at $k=0.7\pi$. This means that, even for the long-range $V$ model, the spectral weight may shift from the $3k_\mathrm{F}$ position to the lower momentum region.
However, a dip at $k=0.6\pi$ remains, inconsistent with the experimental data.

\begin{figure}[tb]
\includegraphics[width=8cm]{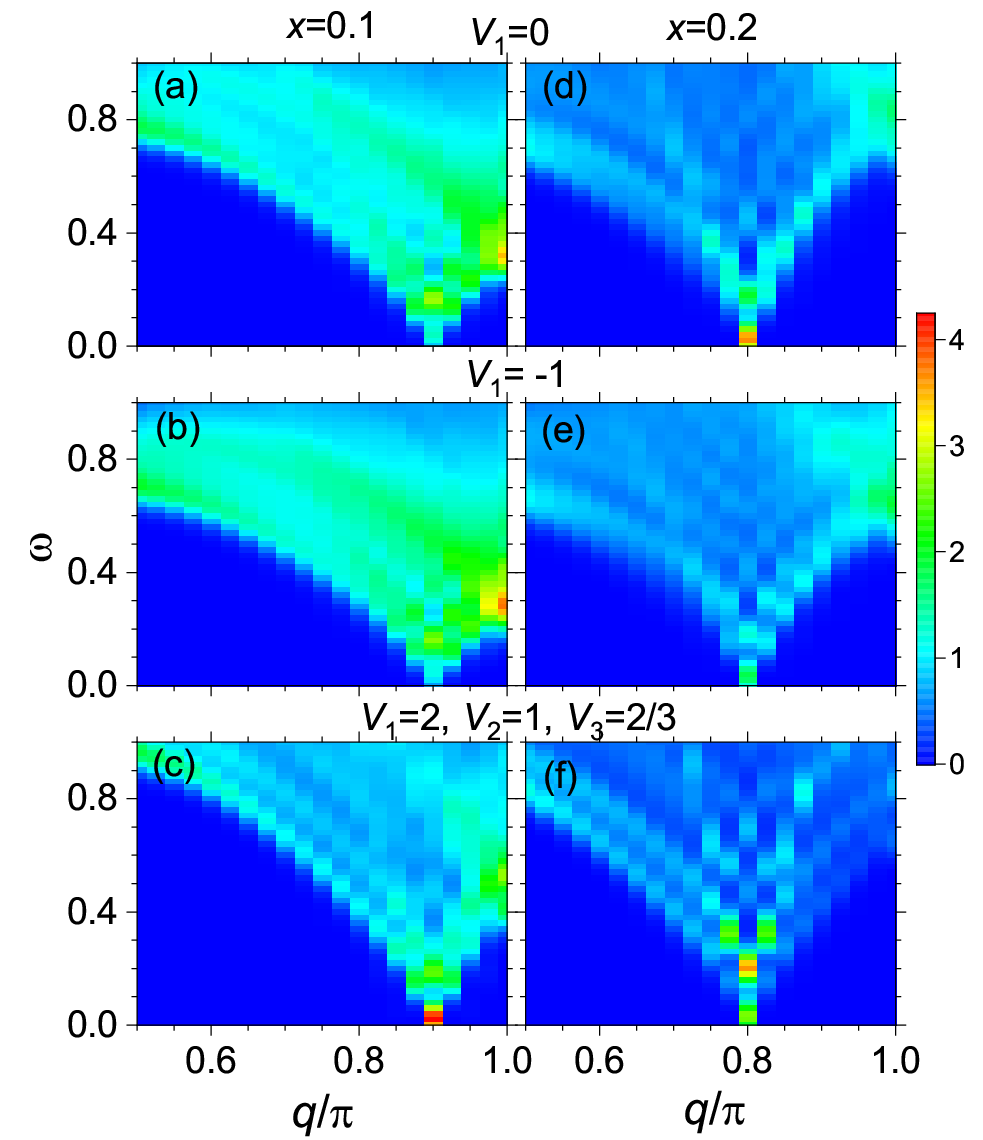}% Here is how to import EPS art
\caption{\label{fig2}Weight map of $S(q,\omega)$ of an $N=80$ Hubbard lattice with $U=8$.  (a) and (d) for the Hubbard model ($V_1=0$), (b) and (e) for the negative $V$ model ($V_1=-1$), and (c) and (f) for the long-range $V$ model ($V_1=2$, $V_2=1$, and $V_3=2/3$) at $x=0.1$ and $x=0.2$, respectively. The scale of spectral weight is shown in the bar at the right-hand side of (e). }
\end{figure}

\subsection{Dynamical spin structure factor}
To clarify how the spin dynamics in doped 1D Mott insulators depends on neighboring interactions, we calculate the dynamical spin structure factor $S(q,\omega)$. Figure~\ref{fig2} shows $S(q,\omega)$ for the three models at $x=0.1$ and 0.2. Since $2k_\mathrm{F}=0.9\pi$ ($0.8\pi$) at $x=0.1$ (0.2), a $2k_\mathrm{F}$ gapless excitation appears at $q=0.9\pi$ ($0.8\pi$) in Fig.~\ref{fig2}(a) [Fig.~\ref{fig2}(d)] as expected. 

The lowest-energy branch of spectral weight corresponds to spinon excitations in the Hubbard model [Figs.~\ref{fig2}(a) and \ref{fig2}(d)]. The energy width of the branch measured from $2k_\mathrm{F}$ is proportional to the nearest-neighbor exchange interaction $J=4t^2/U$ in the large $U$ limit for the Hubbard model. With $V_1$, $J$ is approximately given by $J=4t^2/(U-V_1)$. Therefore, the energy width of the lowest-energy branch is smaller and larger for the negative $V$ model [Figs.~\ref{fig2}(b) and \ref{fig2}(e)] and for the long-range $V$ model [Figs.~\ref{fig2}(c) and \ref{fig2}(f)], respectively, than for the Hubbard model. 

\begin{figure}[tb]
\includegraphics[width=8cm]{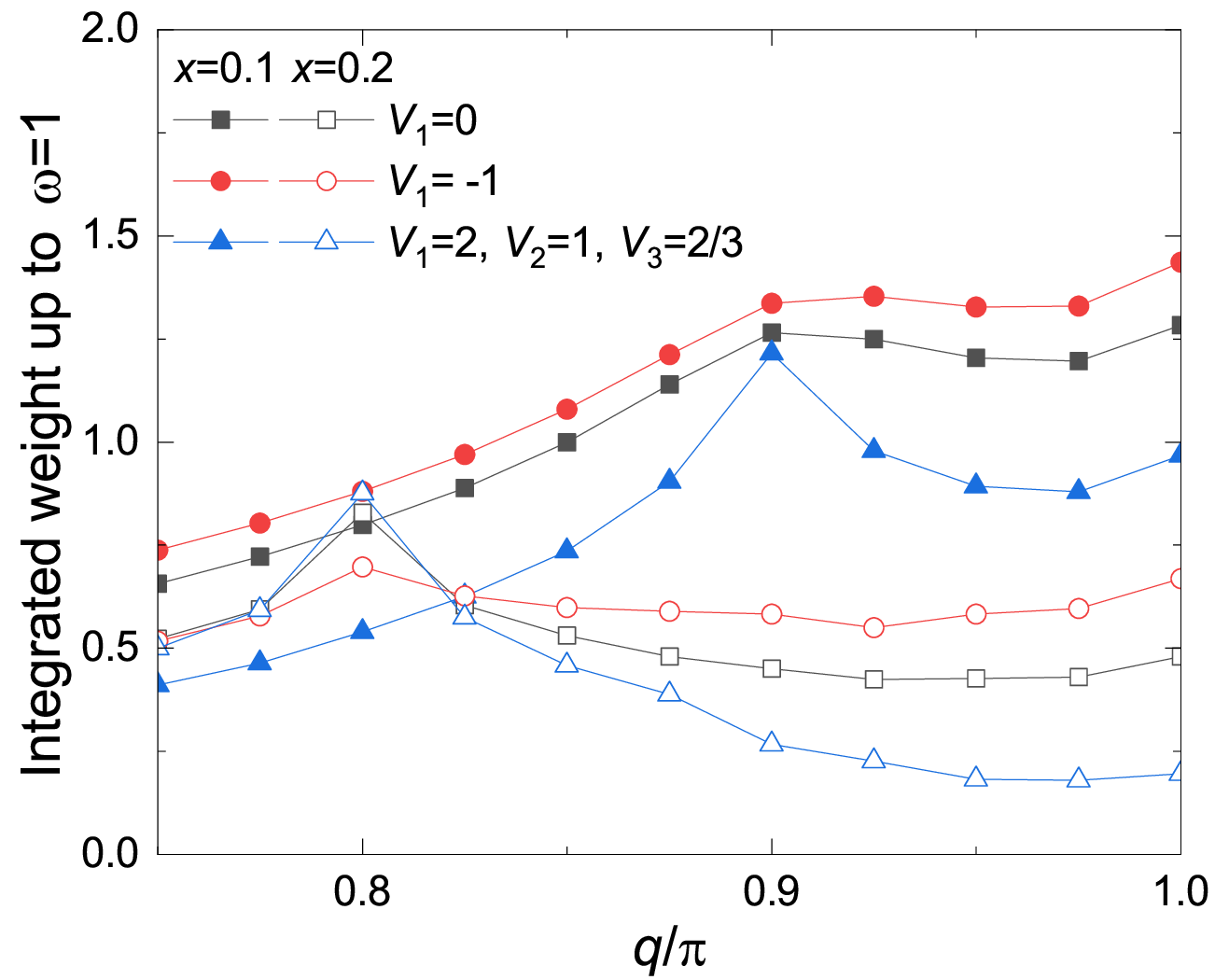}% Here is how to import EPS art
\caption{\label{fig3}Integrated weight of $S(q,\omega)$ up to $\omega=1$ in Fig.~\ref{fig2} for the three models. The solid (open) symbols are for $x=0.1$ ($x=0.2$). }
\end{figure}

The spectral weight distribution in Fig.~\ref{fig2} exhibits a tendency that low-energy $2k_\mathrm{F}$ weight increases with increasing $V_1$ from negative value [Figs.~\ref{fig2}(b) and \ref{fig2}(e)] to positive values [Figs.~\ref{fig2}(c) and \ref{fig2}(f)]. On the other hand, spectral weight around $q=\pi$ is large for the negative $V$ model [Figs.~\ref{fig2}(b) and \ref{fig2}(e)]. To clarify this behavior, we plot the integrated weight of $S(q,\omega)$ up to $\omega=1$ as a function of $q$ in Fig.~\ref{fig3}.  The $q$ dependence of the integrated weights from $2k_\mathrm{F}=0.9\pi$ ($0.8\pi$) for $x=0.1$ (0.2) to $q=\pi$ depends on the models: It is less $q$ dependent for the negative $V$ model than for the long-range $V$ model. Therefore, the $q$ dependence might be a good quantity for judging whether the negative $V$ model or long-range $V$ model is appropriate for the best description of doped 1D cuprates. Since RIXS can observe $q$-dependent spin dynamics, the $q$ dependence would be a possible target for a future RIXS experiment. It is important to notice that the $q$-dependent behavior does not depend on the hole concentration $x$. Therefore, performing analyses on the integrated weight of RIXS for Ba$_{2-x}$Sr$_x$CuO$_{3+\delta}$ with different $x$ will give a possible judgment of the two models.

\begin{figure}[tb]
\includegraphics[width=8cm]{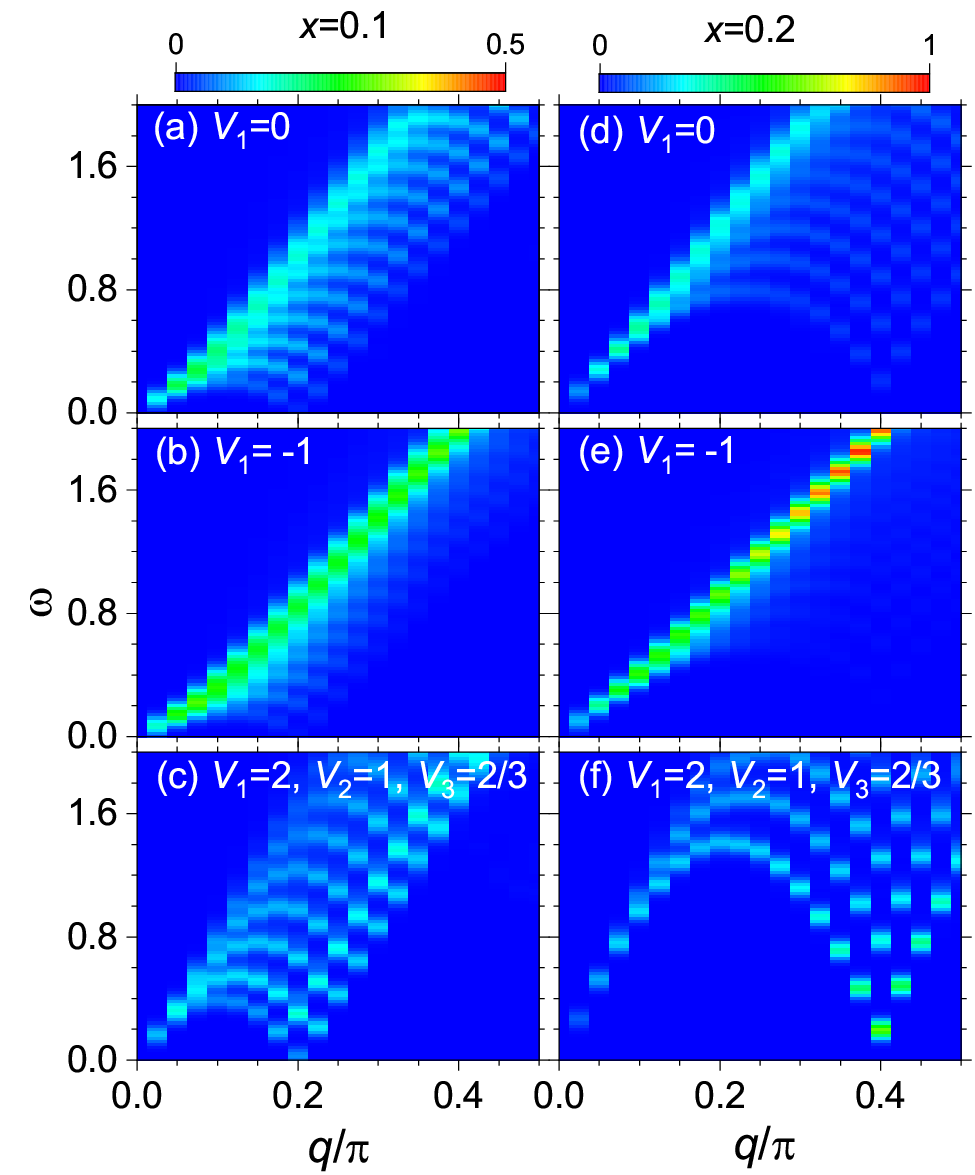}% Here is how to import EPS art
\caption{\label{fig4}Weight map of $N(q,\omega)$ of an $N=80$ Hubbard lattice with $U=8$. (a) and (d) for the Hubbard model ($V_1=0$), (b) and (e) for the negative $V$ model ($V_1=-1$), and (c) and (f) for the long-range $V$ model ($V_1=2$, $V_2=1$, and $V_3=2/3$) at $x=0.1$ and $x=0.2$, respectively. The scale of spectral weight for $x=0.1$ and $x=0.2$ is shown in the bar at the top of (a) and (d), respectively. }
\end{figure}

\subsection{Dynamical charge structure factor}
Charge dynamics also shows contrasting behaviors among the three models. Figure~\ref{fig4} shows $N(q,\omega)$ for the three models at $x=0.1$ and 0.2.  Since $4k_\mathrm{F}=0.2\pi$ ($0.4\pi$) at $x=0.1$ (0.2), a $4k_\mathrm{F}$ gapless charge excitation appears at $q=0.2\pi$ ($0.4\pi$) in Fig.~\ref{fig4}(a) [Fig.~\ref{fig4}(d)] as expected from the 1D Hubbard model. The main weight in the Hubbard model is located at the upper edge of the particle-hole-type excitation in the spectrum. The gradient of the upper edge at $q=0$ corresponds to the charge velocity.

For the long-range $V$ model, spectral weight is uniformly distributed in contrast to the Hubbard model as shown in Figs.~\ref{fig4}(c) and \ref{fig4}(f).  
As a consequence, the spectral weight near the $4k_\mathrm{F}$ gapless point is largest among the three models. If the long-range $V$ model will be applicable to Ba$_{2-x}$Sr$_x$CuO$_{3+\delta}$, the large weight near the $4k_\mathrm{F}$ point may give a large effect on phonon branches near $q=4k_\mathrm{F}$ since the low-energy charge excitation inevitably couples to phonons. In contrast,  such a coupling is expected to be small for the negative $V$ model, where the low-energy excitation around $4k_\mathrm{F}$ is very small as shown in Figs.~\ref{fig4}(b) and \ref{fig4}(e).
We note that the charge velocity is enhanced as compared with that of the Hubbard model,  because of the increase of charge stiffness due to long-range Coulomb interactions. 

In contrast, the spectral weight in the negative $V$ model concentrates highly on the upper edge as shown in Figs.~\ref{fig4}(b) and \ref{fig4}(e). These contrasting behaviors are also a good mark to be judged by RIXS for doped 1D cuprates. 

\begin{figure}[tb]
\includegraphics[width=7cm]{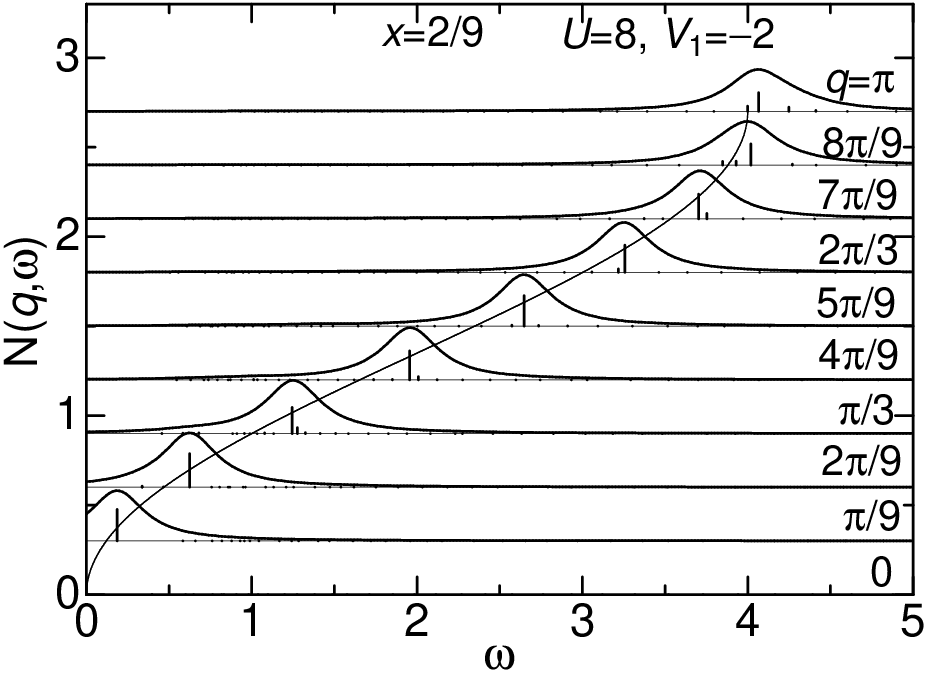}% Here is how to import EPS art
\caption{\label{fig5}$N(q,\omega)$ of an $N=18$ extended Hubbard lattice with $U=8$ and $V_1=-2$ at $x=2/9$. The bars for each momentum represent the weight of the delta function and the solid lines are obtained by broadening the bars with a Lorentzian width of 0.2. The thin sinusoidal curve is $\omega=2(1-\cos q)$ given as a guide for the eyes.}
\end{figure}

To understand the origin of such a large spectral weight at the upper edge for the negative $V$ model, we perform a Lanczos-type exact-diagonalization calculation of $N(q,\omega)$ for a small system size with $N=18$. With increasing absolute value of negative $V_1$, we found that the spectral distribution dramatically changes if $V_1<-2$ (not shown here).  This is due to the presence of a phase-separated state~\cite{Lin1995,Qu2022} for $V_1<-2$. Before the phase separation occurs, we find that the spectral-weight concentration on the upper edge becomes maximum at $V_1=-2$. In Fig.~\ref{fig5}, we show $N(q,\omega)$ at $V_1=-2$ for the $N=18$ lattice, where the spectral weight almost concentrates on a single dispersion. For comparison, a dispersion with $\omega=2(1-\cos q)$ is shown as a thin solid line, which is the same as the 1D ferromagnetic spin-wave dispersion, onto which the charge excitation of a Hubbard model with $U\rightarrow\infty$ and $V_1=-2$ is mapped through the Jordan-Wigner transformation. Therefore, the large spectral weight at the upper edge is a characteristic behavior in the negative $V$ model close to $V_1=-2$.

\section{Summary}
\label{Sec4}
Motivated by the presence of two possible scenarios for understanding electronic states in doped 1D Mott insulators, i.e., the nearest-neighbor attractive interaction and long-range repulsive interaction, we have performed dynamical DMRG calculations of the single-particle spectral function as well as dynamical spin and charge structure factors for the proposed extended Hubbard models. We have found that the spectral function for the attractive interaction captures the ARPES features in Ba$_{2-x}$Sr$_x$CuO$_{3+\delta}$ better than the case of the repulsive interaction. We have also found that the integrated weight of the dynamical spin structure factor up to $\sim 0.6$~eV becomes less momentum-dependent for $q>2k_\mathrm{F}$ and the spectral weight of the dynamical charge structure factor concentrates on the upper edge of the particle-hole-type excitation for the attractive interaction. These behaviors will be confirmed by RIXS if the proposed attractive interaction is dominating the electronic states. Such a confirmation will also justify the reduction of attractive interactions due to the electron-phonon interaction when the system approaches to half-filling~\cite{Wang2021}.

\begin{acknowledgments}
This work was supported by the QST President's Strategic Grant (QST Advanced Study Laboratory); the Japan Society for the Promotion of Science, KAKENHI (Grant No. 19H01829, No. JP19H05825, No. 21H03455, No. 22K03500, No. 24K02948, and No. 24K00560) from the Ministry of Education, Culture, Sports, Science, and Technology, Japan. A part of the computational work was performed using the supercomputing facilities in QST and the computational resources of the supercomputer FUGAKU provided by the RIKEN Center for Computational Science through the HPCI System Research Project (Project ID No. hp230074).
\end{acknowledgments}


\begin{thebibliography}{99}
% ARPES
\bibitem{Chen2021}Z. Chen, Y. Wang, S. N. Rebec, T. Jia, M. Hashimoto, D. Lu, B. Moritz, R. G. Moore, T. P. Devereaux, and Z.-X. Shen, Anomalously strong near-neighbor attraction in doped 1D cuprate chains, Science \textbf{373}, 1235 (2021).
% nondoped ARPES
\bibitem{Kim1996}C. Kim, A. Y. Matsuura, Z.-X. Shen, N. Motoyama, H. Eisaki, S. Uchida, T. Tohyama, and S. Maekawa, Observation of Spin-Charge Separation in One-Dimensional SrCuO$_2$, Phys. Rev. Lett. \textbf{77}, 4054 (1996).
\bibitem{Kim1997}C. Kim, Z.-X. Shen, N. Motoyama, H. Eisaki, S. Uchida, T. Tohyama, and S. Maekawa, Separation of spin and charge excitations in one-dimensional SrCuO$_2$, Phys. Rev. B \textbf{56}, 15589 (1997).
\bibitem{Fujisawa1999}H. Fujisawa, T. Yokoya, T. Takahashi, S. Miyasaka, M. Kibune, and H. Takagi, Angle-resolved photoemission study of Sr$_2$CuO$_3$, Phys. Rev. B \textbf{59}, 7358 (1999).
\bibitem{Kim2006}B. J. Kim, H. Koh, E. Rotenberg, S.-J. Oh, H. Eisaki, N. Motoyama, S. Uchida, T. Tohyama, S. Maekawa, Z.-X. Shen and C. Kim, Distinct spinon and holon dispersions in photoemission spectral functions from one-dimensional SrCuO$_2$, Nat. Phys. \textbf{2}, 397 (2006).
% 1D S-C separation
\bibitem{Essler2005}F. H. L. Essler, H. Frahm, F. G\"{o}hmann, A. Kl\"{u}mper, and V. E. Korepin,  {\it The One-Dimensional Hubbard Model} (Cambridge University Press, Cambridge, U.K., 2005).
% Spectral function for Hubbard model: DDMRG
\bibitem{Benthien2004}H. Benthien, F. Gebhard, and E. Jeckelmann, Spectral Function of the One-Dimensional Hubbard Model away from Half Filling,  Phys. Rev. Lett. \textbf{92}, 256401 (2004).
%Theory ARPES DMRG
\bibitem{Feiguin2024}A. E. Feiguin, C. Helman, and A. A. Aligia, Effective one-band models for the 1D cuprate Ba$_{2-x}$Sr$_x$CuO$_{3+\delta}$, Phys. Rev. B \textbf{108}, 075125 (2023).
% Theory ARPES 
\bibitem{Wang2024}H.-X. Wang, Y.-M. Wu, Y.-F. Jiang, and H. Yao, Spectral properties of a one-dimensional extended Hubbard model from bosonization and time-dependent variational principle: Applications to one-dimensional cuprates, Phys. Rev. B \textbf{109}, 045102 (2024).
% Elec.-phonon
\bibitem{Wang2021}Y. Wang, Z. Chen, T. Shi, B. Moritz, Z.-X. Shen, and T. P. Devereaux, Phonon-Mediated Long-Range Attractive Interaction in One-Dimensional Cuprates, Phys. Rev. Lett. \textbf{127}, 197003 (2021).
% Theory ARPES for elec.-phonon
\bibitem{Tang2023}T. Tang, B. Moritz, C. Peng, Z.-X. Shen, and T. P. Devereaux, Traces of electron-phonon coupling in one-dimensional cuprates, Nat. Commun. \textbf{14}, 3129 (2023).
% Biexciton in 1D Mott
\bibitem{Miyamoto2019}T. Miyamoto, T. Kakizaki, T. Terashige, D. Hata, H. Yamakawa, T. Morimoto, N. Takamura, H. Yada, Y. Takahashi, T. Hasegawa, H. Matsuzaki, T. Tohyama, and H. Okamoto, Biexciton in one-dimensional Mott insulators, Commun. Phys. \textbf{2}, 131 (2019).
% DDMRG
\bibitem{Jeckelmann2002}E. Jeckelmann, Dynamical density-matrix renormalization-group method, Phys. Rev. B \textbf{66}, 045114 (2002).
% DDMRG with polynomial expansion
\bibitem{Sota2010}S. Sota and T. Tohyama, Density matrix renormalization group study of optical conductivity in the one-dimensional Mott insulator Sr$_2$CuO$_3$, Phys. Rev. B \textbf{82}, 195130 (2010).
% Sr2CuO3 optical conductivity
\bibitem{Ono2004}M. Ono, K. Miura, A. Maeda, H. Matsuzaki, H. Kishida, Y. Taguchi, Y. Tokura, M. Yamashita, and H. Okamoto, Linear and nonlinear optical properties of one-dimensional Mott insulators consisting of Ni-halogen chain and CuO-chain compounds, Phys. Rev. B \textbf{70}, 085101 (2004).
% Optical conductivity in 1D extended Hubbard model
\bibitem{Stephan1996}W. Stephan and K. Penc, Dynamical density-density correlations in one-dimensional Mott insulators, Phys. Rev. B {\bf 54}, R17269 (1996).
% Negative V Hubbard
\bibitem{Lin1995}H. Q. Lin, E. R. Gagliano, D. K. Campbell, E. H. Fradkin, and J. E. Gubernatis, The phase diagram of the one-dimensional extended Hubbard model, in {\it The Hubbard model Its Physics and Mathematical Physics}, edited by D. Baeriswyl {\it et al.}, NATO ASI Series B: Physics (Springer, New York, 1995), Vol. 343.
\bibitem{Qu2022}D.-W. Qu, B.-B. Chen, H.-C. Jiang, Y. Wang, and W. Li, Spin-triplet pairing induced by near-neighbor attraction in the extended Hubbard model for cuprate chain, Commun. Phys. \textbf{5}, 257 (2022).

\end{thebibliography}

\end{document}